\documentclass[epj]{svjour}

\usepackage{graphics}
\usepackage{comment}
\usepackage{multirow}
\usepackage{subfigure}
\usepackage{epsfig}

\begin{document}
\title{Link Prediction in Complex Networks: A Clustering Perspective}
\author{Xu Feng, Jichang Zhao \and Ke Xu
\thanks{\emph{Correspondence to:} kexu@nlsde.buaa.edu.cn}%
}                     
%
%
\institute{State Key Laboratory of Software Development Environment, Beihang University, Beijing, 100191, P. R. China}
\date{Received: date / Revised version: date}
%
\abstract{Link prediction is an open problem in the complex network,
which attracts much research interest currently. However, little
attention has been paid to the relation between network structure
and the performance of prediction methods. In order to fill this
vital gap, we try to understand how the network structure affects
the performance of link prediction methods in the view of
clustering. Our experiments on both synthetic and real-world
networks show that as the clustering grows, the precision of these
methods could be improved remarkably, while for the sparse and weakly
clustered network, they perform poorly. We explain this through the
distinguishment caused by increased clustering between the score distribution
of positive and negative instances. Our finding also sheds light on
the problem of how to select appropriate approaches for different
networks with various densities and clusterings.
} 
\maketitle
\section{Introduction}
\label{sec:intro}

Many real world data sets can be represented as networks with nodes
denoting objects and edges describing relationships between them~\cite{Survey}.
Examples of complex networks include the Internet, a collection of
connected Autonomous Systems(AS), routers and interfaces in
different levels. The online social network for people maintaining
their friendship is another major instance. For the pervasive
existence of these networks, the last decade has witnessed the study
of complex networks in the fields of both computer science and physics. An
important issue relevant to the computational analysis of complex
networks is the link prediction. Link prediction is a problem of both
theoretical and practical significance. It aims to evaluate the
likelihood of a link between two nodes not connected until now,
based on the existing links information and possible node attributes
information in the network~\cite{physica}. There are two aspects of link
prediction problem: on the one hand, for most real network data, not
all links are already observed, link prediction helps to find the
missing links; on the other hand, it can help us infer the new
interactions between nodes in the new future. Research on link
prediction is also helpful to accomplish some other tasks, like
collective classification~\cite{CCLP} and anomalous link discovery~\cite{ALD}.

The existing methods for link prediction can be divided into three
categories. The first method defines a measure of proximity or
similarity between two nodes in the network, taking into account
that links between more similar nodes are of higher existing
likelihood. Liben-Nowell and Kleinberg~\cite{Nowell} summarize many
similarity measures based on node neighborhoods, the ensemble of all
paths and higher-level approaches. They compare these measures with
random predictors in five co-authorship networks and find that there
is indeed useful information contained in the network topology
alone. Motivated by the resource allocation process taking place in
networks, Zhou et al.~\cite{RA} propose a new similarity measure,
which has great performance in six representative networks drawn from
different fields. Liu and L\"{u}~\cite{SRW} put forward a method
based on local random walk, which can give excellent prediction
while has low computational complexity. The second method is based
on the maximum likelihood estimation. Empirical studies suggest that
many real-world networks exhibit hierarchical organization. Clauset,
Moore and Newman~\cite{Nature} present a method inferring
hierarchical structure from network data and use the knowledge of
hierarchical structure to predict the missing links in partially
known networks. The third method mainly uses machine learning
techniques. Hasan et al~\cite{Supervised_Learning} view link
prediction as a supervised learning task: for two potentially
connected nodes, predicting whether it is a positive or negative
example. The feature set extracted from the co-authorship graph
contains proximity features, aggregated features and topological
features. They experiment with seven different classification
algorithms and compare the performance of these classifiers using
different performance metrics. O'Madadhain et
al.~\cite{Event_network} use primarily probabilistic classifiers to
predict future \lq\lq co-participating\rq\rq~in event-based network
data. There are also many works related to link prediction
concerning more complicated networks, like directed and weighted
networks. Leung et al.~\cite{LFR} propose a novel Link
Formation Rules mining algorithm for social networks. Romero and
Kleinberg~\cite{Closure} investigate the directed closure process
and analyze the link formation on twitter.
In~\cite{weight_network}, Murata and Moriyasu describe an improved
method for predicting links on Question-Answering Bulletin Boards
(QABB), kind of a social network in which each link is assigned a
weight.

Most of those works on link prediction aim to find a method with
better prediction performance for some particular networks, such as
the co-authorship network, terrorists network and so on. However,
little was done to reveal how these existing methods perform on
networks with different structural properties. In this paper, we try
to find the relation between network structure and the prediction
performance of these methods. In the real world, the attributes of
nodes are usually difficult to collect and the simpleness of prediction
methods is also necessary. For example, in online social
networks, systems need to provide a list of potential friends for
a certain user with least load to the server. Because of this, in
the present work, we focus on the first kind of methods which are
solely based on the network structure. Through experiments on both
synthetic and real-world networks, we find that for the network with
low clustering, these methods perform poorly. Nonetheless, as
the clustering of the network grows, the precision of these methods
is drastically improved. These phenomena tell us that for the
networks with various clusterings, we should employ different
methods for link prediction.

This rest of the present paper is organized as follows: In
Section~\ref{sec:preliminaries}, we review several
similarity based methods for link prediction. In
Section~\ref{sec:datasets}, the data sets we use in the paper are
introduced. We investigate the connection between clustering and
performance of prediction methods in Section~\ref{sec:experiments},
and we also give a brief explanation in this section. In
Section~\ref{sec:conslusion}, we conclude this work briefly.

\section{Preliminaries}
\label{sec:preliminaries}

In this section, we first describe the link prediction problem and
introduce the evaluation metrics. Then we review several
similarity-based methods.

Suppose we have an undirected simple network $G(V, E)$, where $V$ is
the set of nodes and $E$ is the set of edges. Generally, the number
of a node's connections can be defined as its degree. The averaged
degree of the network can be defined as $$\langle k
\rangle=\frac{2|E|}{|V|},$$ which could be used to characterize the
density of the network. We use $p(k)$ to denote the degree distribution of the
network and for the complex networks discussed in this paper, it
always follows a power-law. The relative size of the giant connected
component($GCC$) can be denoted as $f_{GCC}.$ Clustering of a node $i$
is used to characterize how closely its neighbors are connected. It
can be defined as
$$C_i=\frac{2|E_i|}{k_i(k_i-1)},$$ where $E_i$ is the set of ties
between $i$'s neighbors and $k_i$ is the degree of $i$. We do not
take the case of $k_i=1$ into consideration. The averaged clustering
of the network can be defined as
\begin{equation}
C=\frac{\sum_{\{i \in V\}}{C_i}}{|V|}.
\end{equation}
In the rest of paper, we omit the word \lq\lq averaged\rq\rq~if
there is no confusion in the context.

For any pair of nodes $\langle x, y\rangle$, which is not existing in
$E$, each similarity-based method defines a measure, i.e. a score
$s(x, y)$ is assigned according to the given network topology. Then
we rank all of these scores of node pairs and a higher score means a
higher probability that the corresponding link will emerge in the
future or more likely be missed in the present sample.

To test the prediction accuracy of each method, we adopt the approach
used in~\cite{RA}. The edge set $E$ is randomly divided into two
parts, including $E^{Train}$ and $E^{Test}$, respectively. The
training set $E^{Train}$ is supposed to be known information and
$E^{Test}$ is the testing set consisting of missing links or links
to occur in the future. The training set contains 90\% of
links in $E$, and the remaining 10\% of links are in the testing
set. We use precision to quantify the accuracy of prediction
measures, which is determined as follows. Let $n$ denote the number of
links in $E^{Test}$. We compute the score list based on $G(V,
E^{Train})$ and rank the list in decreasing order. The first $n$
pairs are taken and $m$ denotes the size of the intersection of this
set of pairs with the $E^{Test}$. Then the precision is $P=m/n$.

We mainly explore six existing similarity-based measures for link
prediction, including (1) Common Neighbors(\texttt{CN}); (2)
Adamic-Adar Index(\texttt{AA}); (3) Resource Allocation
Index(\texttt{RA}); (4) Katz Index(\texttt{Katz}); (5) Rooted
PageRank(\texttt{PR}); (6) Superposed Random Walk(\texttt{SRW}). A
brief introduction of these methods is given as follows.

\textbf{Common Neighbors} For a node $x$ in $G$, $N(x)$ denotes the
set of neighbors of $x$. The \emph{Common Neighbors} measure is
determined by the number of nodes that link to both $x$ and $y$,
that is to say, two nodes is more likely to be connected with more
common neighbors. Therefore, the score can be defined as
\begin{equation}
s^{CN}(x,y)=|N(x)\cap N(y)|.
\end{equation}

\textbf{Adamic-Adar Index} In~\cite{AA}, to determine whether two
personal home pages are strongly \lq\lq related\rq\rq, Adamic and
Adar define the similarity between two pages based on their shared
features. For link prediction, this index assigns rarer connected
node more weights, i.e.,
\begin{equation}
s^{AA}(x,y)=\sum_{z \in N(x)\cap N(y)}{1/log(k(z))},
\end{equation}
where $k(z)$ is the degree of the node $z$.

\textbf{Resource Allocation Index} Zhou et al.~\cite{RA} consider such a
process: for a pair of unconnected nodes $x$ and $y$, $x$ with a
unit of resource can send some to $y$ by sending averaged amounts to
its neighbors. The more resource $y$ receives from $x$, the more
likely a link between $x$ and $y$ exists. Therefore, the score
between x and y is defined as
\begin{equation}
s^{RA}(x,y)=\sum_{z \in N(x) \cap N(y)}{1/k(z)}.
\end{equation}

\textbf{Katz Index} \emph{Katz Index} is a path-ensemble based
method. It sums over all paths between $x$ and $y$. The more number
of paths with short length, the higher the score is. It is defined
as
\begin{equation}
s^{Katz}(x,y)=\sum_{l=1}^{\infty}{\beta^l \cdot |paths^{l}_{x,y}|},
\end{equation}
where $\beta$ is an adjusting parameter and $paths^{l}_{x,y}$ is the
set of all paths with length $l$ from $x$ to $y$. As mentioned in
~\cite{Nowell}, we can get the score matrix $S^{Katz}$ by
\begin{equation}
S^{Katz}=\sum_{l=1}^{\infty}{\beta ^{l} A^l}=(I-\beta A)^{-1}-I,
\end{equation}
where $I$ is the identity matrix and $A$ is the adjacent matrix of $G$.

\textbf{Rooted PageRank Index} The \emph{Rooted PageRank} defines a
random walk on the underlying graph $G$. A random walk starts from a
node $x$, and iteratively moves to a neighbor of $x$ chosen
uniformly at random. We use the probability that a random walk
starting from $x$ runs into $y$ as the indicator of similarity
between $x$ and $y$~\cite{imc}. The $s^{PR}(x, y)$ under the \emph{Rooted
PageRank} is defined to be the stationary probability of $y$ under
such a random walk: with probability $1-\beta$ returns to $x$ at
each step, moves to a random neighbor of the current node with
probability $\beta$. Let $$D_{ii}=1/\sum{A_{ij}},$$  $D_{ij}=0$ when
$i\neq j$ and
$$T=DA,$$ we have the score matrix $S^{PR}$
\begin{equation}
S^{PR}=(1-\beta)(I-\beta T)^{-1}.
\end{equation}

\textbf{Superposed Random Walk} Liu and L\"{u}~\cite{SRW} propose
the \emph{Superposed Random Walk Index}, which focuses on just
few-step random walk, rather than the stationary probability. The
transition probability matrix is denoted as $P$, with
$P_{xy}=a_{xy}/k_{x},$ where $a_{xy}$ represents the corresponding
entry in $A$. Given a random walk starting at $x$, the probability
that it locates at $y$ after $t$ steps is $\pi_{xy}(t)$.
$\overrightarrow{\pi}_{x}(0)$ is a $N*1$ vector with $x^{th}$
element equals 1 and others equal 0. Then we have:
\begin{equation}
\overrightarrow{\pi}_{x}(t)=P^{T}\overrightarrow{\pi}_{x}(t-1).
\end{equation}

The similarity based on \emph{Local Random Walk} is defined as:
\begin{equation}
s^{LRW}_{xy}(t)=\frac{k_{x}}{2|E|}\cdot\pi_{xy}(t)+\frac{k_{y}}{2|E|}\cdot\pi_{yx}(t).
\end{equation}
The \emph{Superposed Random Walk Index} superposes
the contribution of independently moved walkers and in our configuration
we compute the \textbf{3} steps \texttt{SRW} rather than the
optimal-steps \texttt{SRW}. The score for the pair $\langle
x,y\rangle$ can be defined as
\begin{equation}
s^{SRW}(x,y,t)=\sum_{\tau=1}^{t}{s_{xy}^{LRW}(\tau)}.
\end{equation}

Through the measure of precision and these prediction
methods, we then perform experiments on both the synthetic and
real-world networks that will be introduced in the next section.

\section{Data Sets}
\label{sec:datasets}

The complex networks are pervasively existing in the real-world.
Empirical study suggests that most complex networks exhibit the
\lq\lq scale-free\rq\rq~property, which means $p(k) \propto
k^{-\gamma}$. Barab\'{a}si and Albert~\cite{BA} proposed a
scale-free network model to explain the generation mechanism of the
\lq\lq power-law\rq\rq~distribution, known as the BA model. We utilize
this model to generate the synthetic networks. We denote the network
generated by the BA model as $BA(N, m)$, where $N$ is the size of
the network generated, $m$ is the number of links that a new node
will establish when it is added to the network and the
averaged degree is $2m$. We generate five networks in this paper.

We also import three typical real-world complex networks collected
from different fields. \texttt{Netscience} is a network of
co-authorships between scientists who are themselves publishing on
the topic of network science~\cite{netscience}. There are 1589
scientists in this network and 128 of them are isolated. We will not use these isolated nodes in our experiment. \texttt{Power
Grid} is a well-connected electrical power grid of western US,
where nodes denote generators, transformers and substations and edges
denote the transmission lines between them~\cite{Grid}. \texttt{Politic Blog} is
a directed network of US political blogs~\cite{PB}. Here we treat
its links as undirected and self-connections are omitted.

The detailed descriptions of these data sets are listed in
Table~\ref{tab:ba-datasets}.
\begin{table}
\centering \caption{Synthetic and real-world data sets}
\label{tab:ba-datasets}
\begin{tabular}{llllll}
\hline\noalign{\smallskip}
Network & $|V|$ & $|E|$ & $\langle k \rangle$ & $C$ & $f_{GCC}$\\
\noalign{\smallskip}\hline\noalign{\smallskip}
\texttt{BA(1000,2)} & 1000 & 1997 & 4 & 0.027 & 1\\
\texttt{BA(1000,5)} & 1000 & 4985 & 10 & 0.039 & 1\\
\texttt{BA(1000,10)} & 1000 & 9945 & 20 & 0.064 & 1\\
\texttt{BA(2000,5)} & 2000 & 9985 & 10 & 0.024 & 1\\
\texttt{BA(4000,5)} & 4000 & 19985 & 10 & 0.017 & 1\\
\texttt{Netscience} & 1461 & 2742 & 3.75 & 0.878 & 0.26\\
\texttt{Power Grid} & 4941 & 6594 & 2.67 & 0.107 & 1\\
\texttt{Politic Blog} & 1224 & 16715 & 27.31 & 0.36 & 0.998\\
\noalign{\smallskip}\hline
\end{tabular}
\end{table}

\section{How Clustering affects Predicting Precision}
\label{sec:experiments}

In this section, we first perform experiments on synthetic
networks with various clusterings and unveil the relation between the
network structure and the precision of link prediction methods. Then
we validate our findings on the representative real-world data sets.
Finally, we give an explanation based on class distribution for the phenomenon.

\subsection{Results from Synthetic Networks}
\label{subsec:resultsfromba}

We investigate the relationship between the clustering and the
precision of link prediction in this subsection. In order to unveil this relationship,
the variation of clustering of the network is necessary. For this
reason, we use the method proposed by Kim et
al.~\cite{kim_clustering} to rewire the links randomly and achieve
the purpose of varying the clustering but without changing the
degrees of the nodes. In particular, we first randomly pick up two
edges, say $\langle A,B\rangle$ and $\langle C,D\rangle$. We then
compare the numbers of local triangular structures associated with
all three configurations $\{\langle A,B\rangle, \langle
C,D\rangle\}$, $\{\langle A,C\rangle, \langle B,D\rangle\}$ and
$\{\langle A,D\rangle, \langle B,C\rangle\}$, and select the one
with most triangles and connect the nodes accordingly, where
duplicated links are avoided~\cite{ma_clustering}. It is worthy to
note that, in this process, if a link from a given node is detached,
a different link is immediately attached to this node. We continue
this process until a desirable value of $C$ of the network is
attained. In this approach, the degree sequence of the network are
fixed, and the only topological property changed is $C$.

\begin{figure}
\resizebox{0.53\textwidth}{!}{%
  \includegraphics{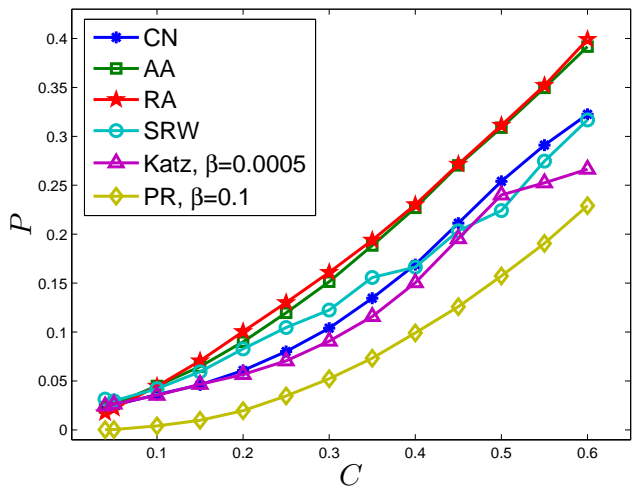}
}
\caption{Link prediction on \texttt{BA(1000,5)} with varying
clustering}
\label{fig:methods_ba_1000_5}       
\end{figure}

\begin{table*}
\centering \caption{The result from \texttt{BA(4000,5)}}
\label{tab:ba_4000_results}
\begin{tabular}{lllllllllll}
\hline\noalign{\smallskip}
\multirow{2}{*}{$C$} & \multirow{2}{*}{\texttt{CN}} & \multirow{2}{*}{\texttt{AA}} & \multirow{2}{*}{\texttt{RA}} & \multirow{2}{*}{\texttt{SRW}} & \multicolumn{3}{|c|}{\texttt{Katz}} & \multicolumn{3}{|c|}{\texttt{PR}}\\
\noalign{\smallskip}\hline\noalign{\smallskip} & & & & &
$\beta=0.05$ &
$\beta=0.005$ & $\beta=0.0005$ & $\beta=0.1$ & $\beta=0.5$ & $\beta=0.9$\\
\hline
0.0176 & 0.0134 & 0.0128 & 0.0096 & 0.0145 & 0.0003 & 0.0134 & 0.0150 & 9.91E-05 & 5.04E-05 & 0.0012\\
0.1 & 0.0393 & 0.0595 & 0.0541 & 0.0509 & 0.0159 & 0.0318 & 0.0305 & 0.0046 & 0.0038 & 0.0051\\
0.15 & 0.0557 & 0.0901 & 0.0846 & 0.0704 & 0.0321 & 0.0404 & 0.0383 & 0.0091 & 0.0095 & 0.0105\\
0.2 & 0.0722 & 0.1189 & 0.1159 & 0.0932 & 0.0487 & 0.0534 & 0.0544 & 0.0205 & 0.0183 & 0.0185\\
0.25 & 0.0925 & 0.1461 & 0.1430 & 0.1132 & 0.0592 & 0.0638 & 0.0657 & 0.0358 & 0.0334 & 0.0326\\
0.3 & 0.1168 & 0.1781 & 0.1726 & 0.1342 & 0.0616 & 0.0832 & 0.0800 & 0.0522 & 0.0532 & 0.0501\\
0.35 & 0.1432 & 0.2130 & 0.2037 & 0.1610 & 0.0937 & 0.1054 & 0.1067 & 0.0744 & 0.0702 & 0.0690\\
0.4 & 0.1785 & 0.2514 & 0.2405 & 0.1827 & 0.1060 & 0.1383 & 0.1410 & 0.0979 & 0.0943 & 0.0898\\
0.45 & 0.2162 & 0.2898 & 0.2774 & 0.2140 & 0.1203 & 0.1790 & 0.1841 & 0.1184 & 0.1184 & 0.1104\\
0.5 & 0.2573 & 0.3279 & 0.3163 & 0.2458 & 0.1320 & 0.2391 & 0.2346 & 0.1541 & 0.1519 & 0.1364\\
0.55 & 0.3074 & 0.3704 & 0.3616 & 0.2824 & 0.1152 & 0.2983 & 0.3025 & 0.1882 & 0.1853 & 0.1702\\
0.6 & 0.3450 & 0.4132 & 0.4081 & 0.3154 & 0.1241 & 0.3226 & 0.3292 & 0.2267 & 0.2143 & 0.1903\\
\noalign{\smallskip}\hline
\end{tabular}
\end{table*}

We perform the six methods on different generated networks with
various clusterings. The results from \texttt{BA(1000,5)} are shown
in Fig.~\ref{fig:methods_ba_1000_5}. As for \texttt{Katz} and
\texttt{PR}, we only choose the best situation to represent
them, i.e. $\beta=0.0005$ for \texttt{Katz} and $\beta=0.1$ for
\texttt{PR}. It can be seen that while the clustering $C$ increases,
all these similarity-based methods have better prediction
performance. In networks with relatively small $C$, e.g. $C<0.01$,
there doesn't exist a more competitive method, and while
$C$ grows to a certain value, \texttt{AA} and \texttt{RA} seem to be
better. The results from other synthetic data sets are similar. For
example, Table~\ref{tab:ba_4000_results} shows the
result from \texttt{BA(4000,5)}. We can see that the correlated
characteristic between $C$ and prediction value of different methods
does not vary with the size and density of the network. It is also
interesting that for the method of \texttt{Katz}, its performance
depends on the value of $\beta$ greatly. For instance, when
$\beta=0.0005$, it performs best as clustering grows. This
phenomenon means that the nearest neighbors play a vital role in the
prediction for \texttt{Katz}, however, considering the further hops
is unnecessary.

\begin{figure*}
\centering
 \subfigure[\scriptsize $C=0.1$]{\epsfig{file=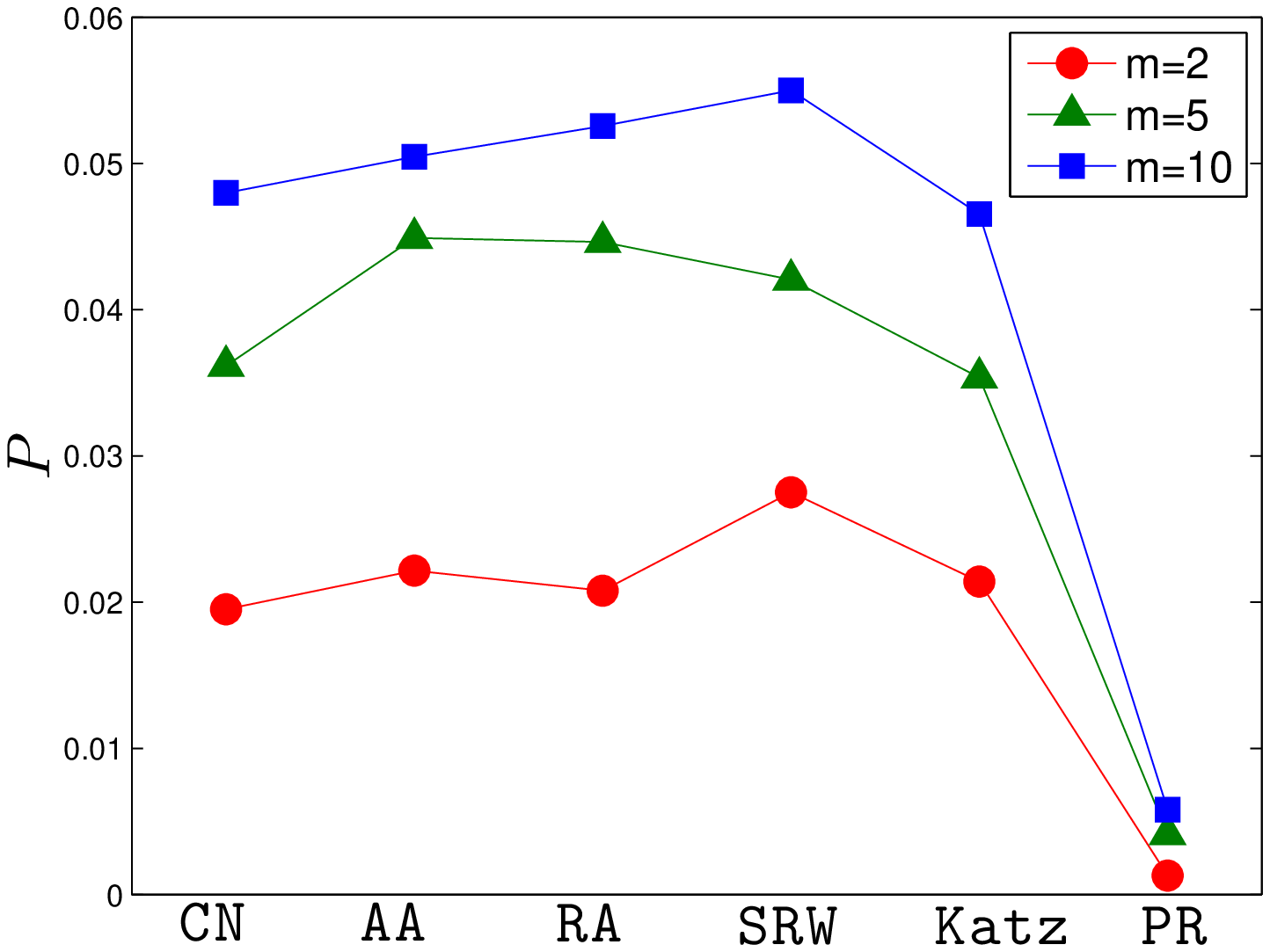,width=5.8cm}\label{fig:ba_c1m}}
 \subfigure[\scriptsize $C=0.3$]{\epsfig{file=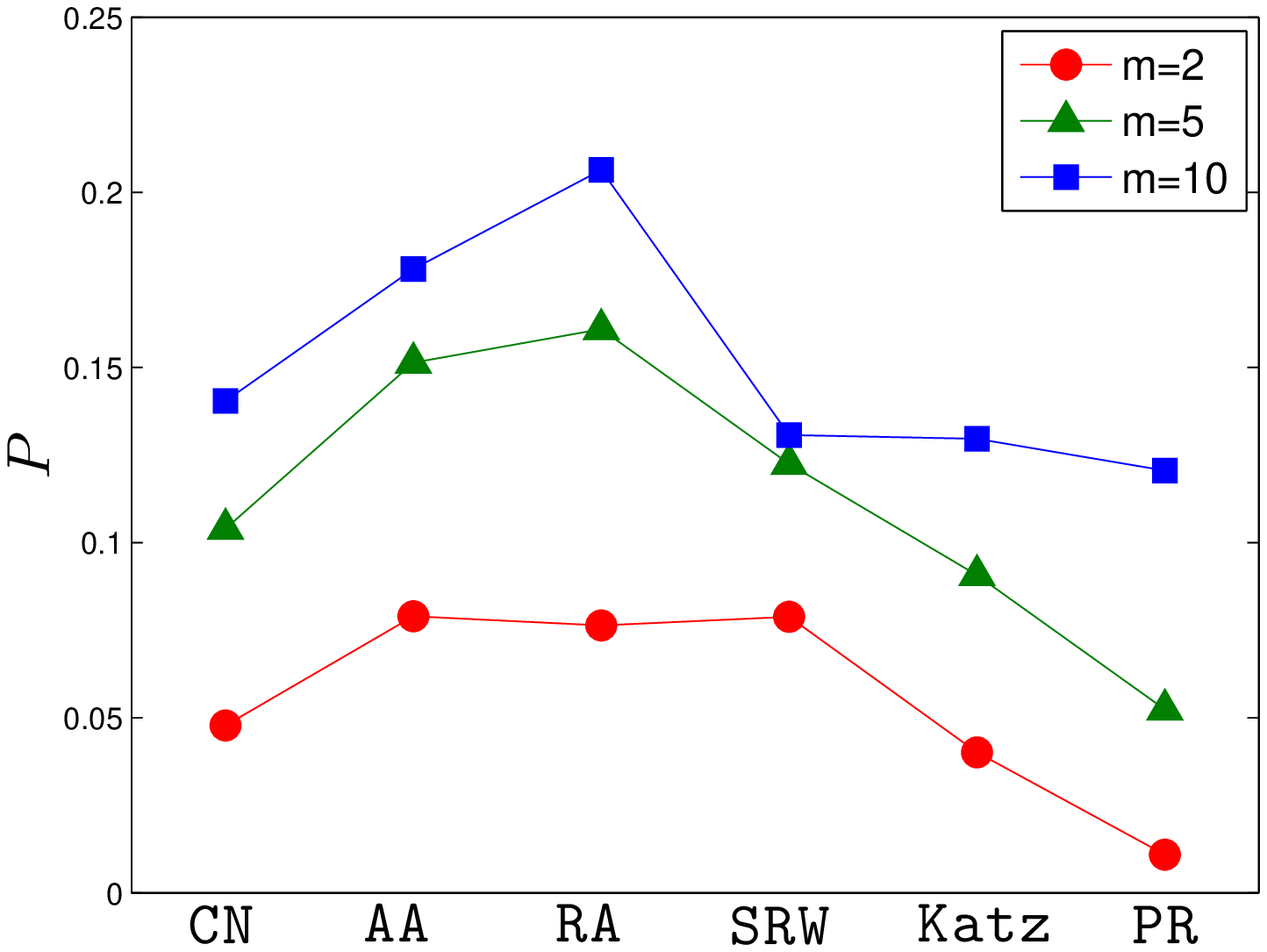,width=5.8cm}\label{fig:ba_c3m}}
 \subfigure[\scriptsize $C=0.5$]{\epsfig{file=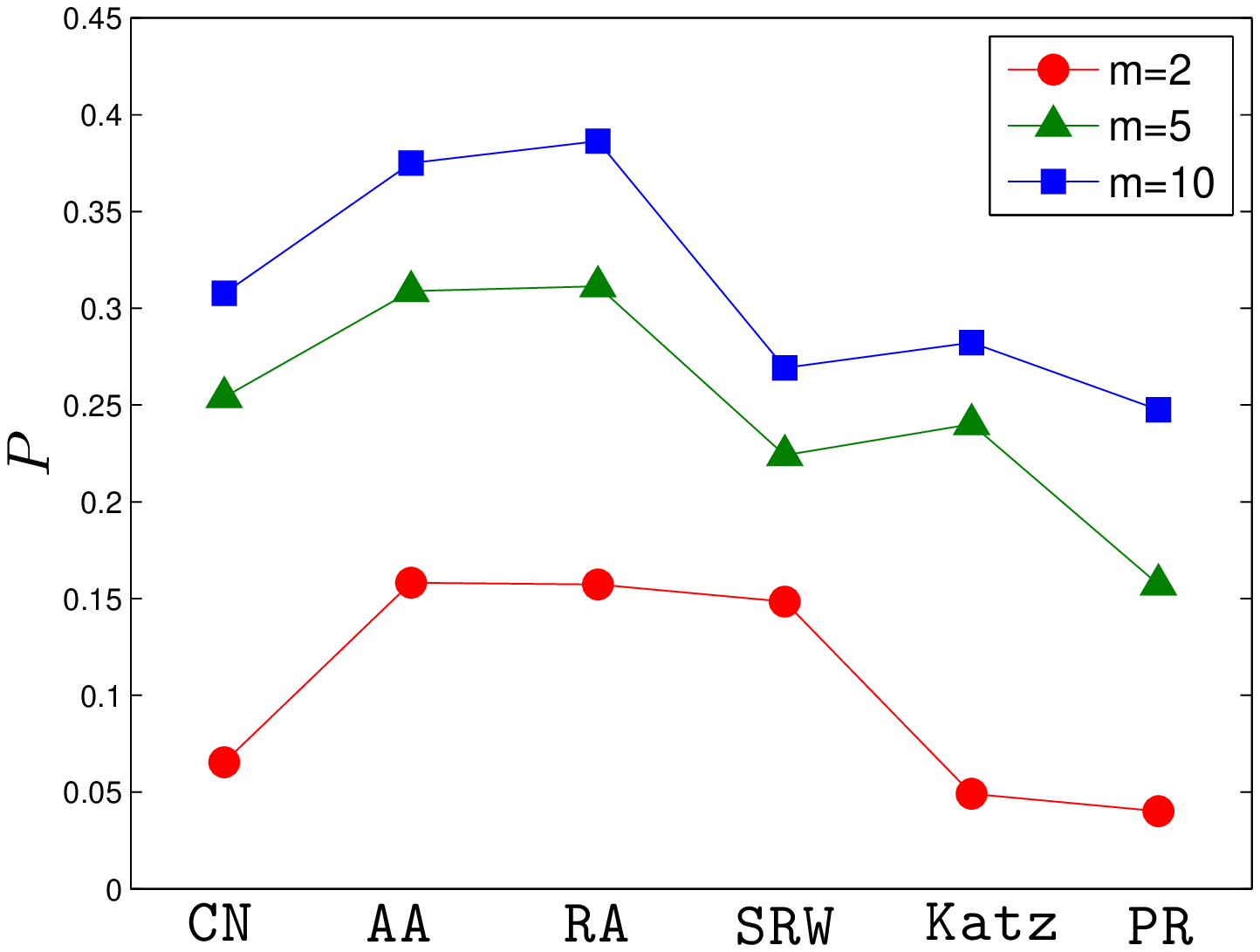,width=5.8cm}\label{fig:ba_c5m}}
\caption{$P$ varies with density of the network}
\label{fig:pvsdensity}
\end{figure*}

Meanwhile, as shown in Fig.~\ref{fig:pvsdensity}, we choose three
representative clusterings, i.e., $C=0.1$, $C=0.3$ and
$C=0.5$, to observe how these methods perform on networks with
different densities when the size and clustering of the networks are
constant. It is easy to learn from Fig.~\ref{fig:pvsdensity} that
these methods perform better on denser networks. However, for the
sparse network with low clustering, say $C=0.1$, as shown in
Fig.~\ref{fig:ba_c1m}, \texttt{SRW} performs best compared with
other approaches when $m=2$. Nevertheless, the situation changes
when the clustering of the network grows, \texttt{AA} and
\texttt{RA} perform better, too. In particular, \texttt{RA} is the
best way among these methods for the dense network with high
clustering, as shown in Fig.~\ref{fig:ba_c5m}.

In summary, we find that on the synthetic networks generated by BA
model, when the clustering grows, the performance of these
prediction methods improves. However, a natural question is
whether similar phenomenon can be found in real-world networks.
Therefore, we validate this finding on the real-world
data sets in the next subsection.

\subsection{Validation on Real-world Data Sets}
\label{subsec:validation}

The result of link prediction experiments on these real-world
networks is shown in Table~\ref{tab:realworlddatasets}, which
is consistent with the above simulation experiment. We can see that
the prediction methods perform best on \texttt{Netscience} which has
the largest clustering and worst on \texttt{Power Grid} with the
least $C$.

\begin{table*}
\centering \caption{The result from real-world data sets}
\label{tab:realworlddatasets}
\begin{tabular}{lllllllllll}
\hline\noalign{\smallskip}
\multirow{2}{*}{\texttt{Network}} & \multirow{2}{*}{\texttt{CN}} & \multirow{2}{*}{\texttt{AA}} & \multirow{2}{*}{\texttt{RA}} & \multirow{2}{*}{\texttt{SRW}} & \multicolumn{3}{|c|}{\texttt{Katz}} & \multicolumn{3}{|c|}{\texttt{PR}}\\
\noalign{\smallskip}\hline\noalign{\smallskip} & & & & &
$\beta=0.05$ &
$\beta=0.005$ & $\beta=0.0005$ & $\beta=0.1$ & $\beta=0.5$ & $\beta=0.9$\\
\hline
\texttt{Netscience} &    0.4494 &    0.6666 &    \textbf{0.6805} &    0.5760 &    0.3796 &    0.4569 &    0.4423 &    0.3734 &    0.3817 &    0.3013\\
\texttt{Power Grid} &     \textbf{0.0438} &    0.0281 &    0.0252 &    0.0227 &    0.0085 &    0.0067 &    0.0110 &    0.0085 &    0.0067 &    0.0110\\
\texttt{Politic Blog} &  0.1724 &    0.1712 &    0.1497 &    0.1421 &    0.0309 &    \textbf{0.1776} &    0.1733 &    0.0141 &    0.0288 &    0.0537\\
\noalign{\smallskip}\hline
\end{tabular}
\end{table*}

Based on the validations above, we can conjecture that in
real-world networks, the performance of these link prediction methods
is closely related to their clusterings. That is, for the network with
higher clustering, these methods perform better. However, when the
clustering decreases, their precision drops.

\subsection{An Explanation Based on Class Distribution}
\label{subsec:explanaton}

In this subsection, we try to explain the finding in the view of
class distribution. Here we treat the pair of connected nodes as a
positive instance while the pair of disconnected nodes is a negative
instance. As mentioned in~\cite{ALD}, the highly skewed distribution
of positive and negative examples yields computational cost of all
node pairs and increases the variance of the prediction model. We
assume that the scores of each particular link prediction method are
drawn from separate distributions for linked and non-linked node
pairs. In principle, the similarity-based method for link prediction
tries to distinguish the two distributions of positive and negative
examples by the corresponding scores. Next, we focus on how the
distribution of scores on two types of node pairs varies as the
network structure changes.

\begin{figure*}
\centering
\centering
 \subfigure[\scriptsize \texttt{CN}, $C=0.039$]{\epsfig{file=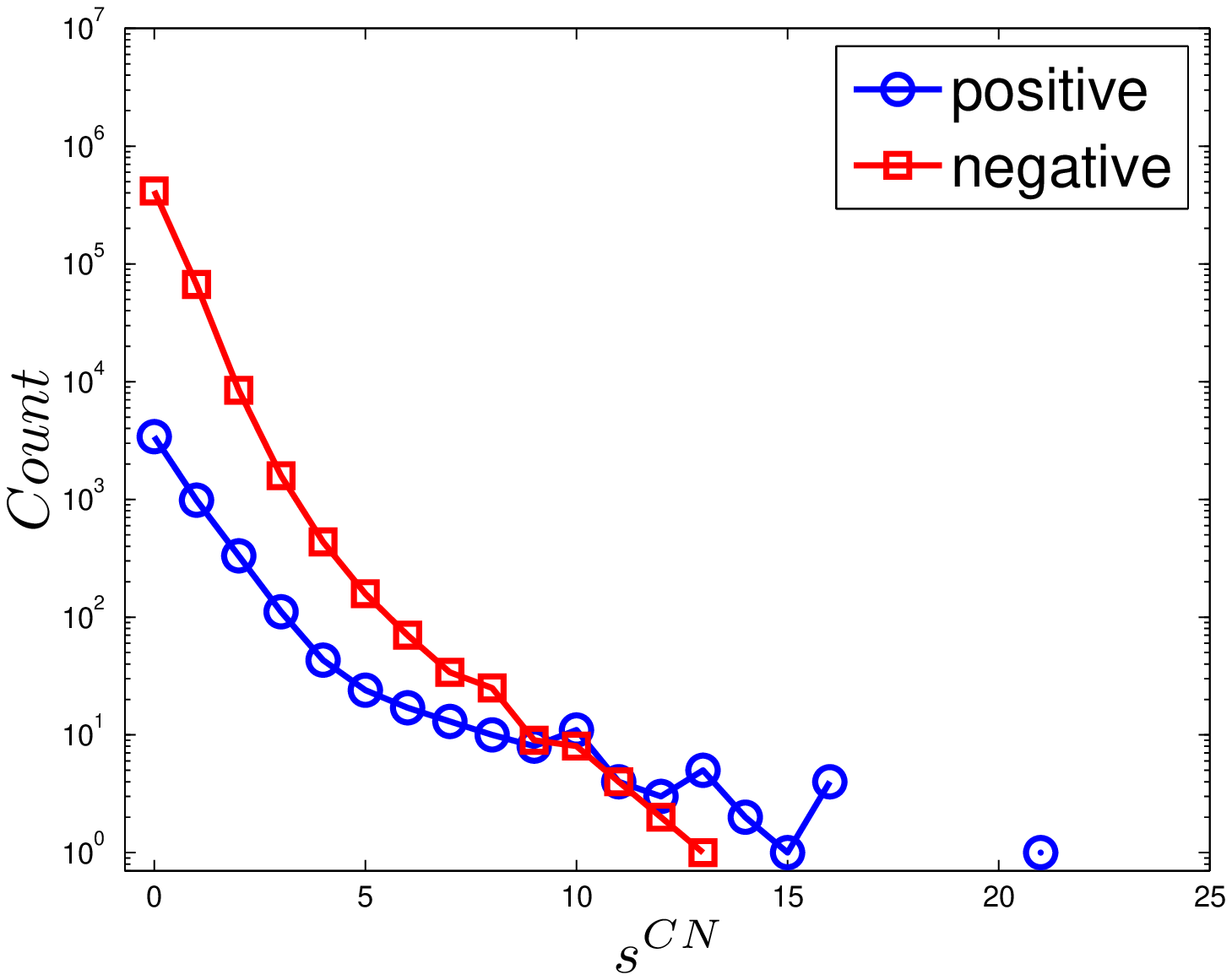,width=5.8cm}\label{fig:ba_cn_origin}}
 \subfigure[\scriptsize \texttt{CN}, $C=0.3$]{\epsfig{file=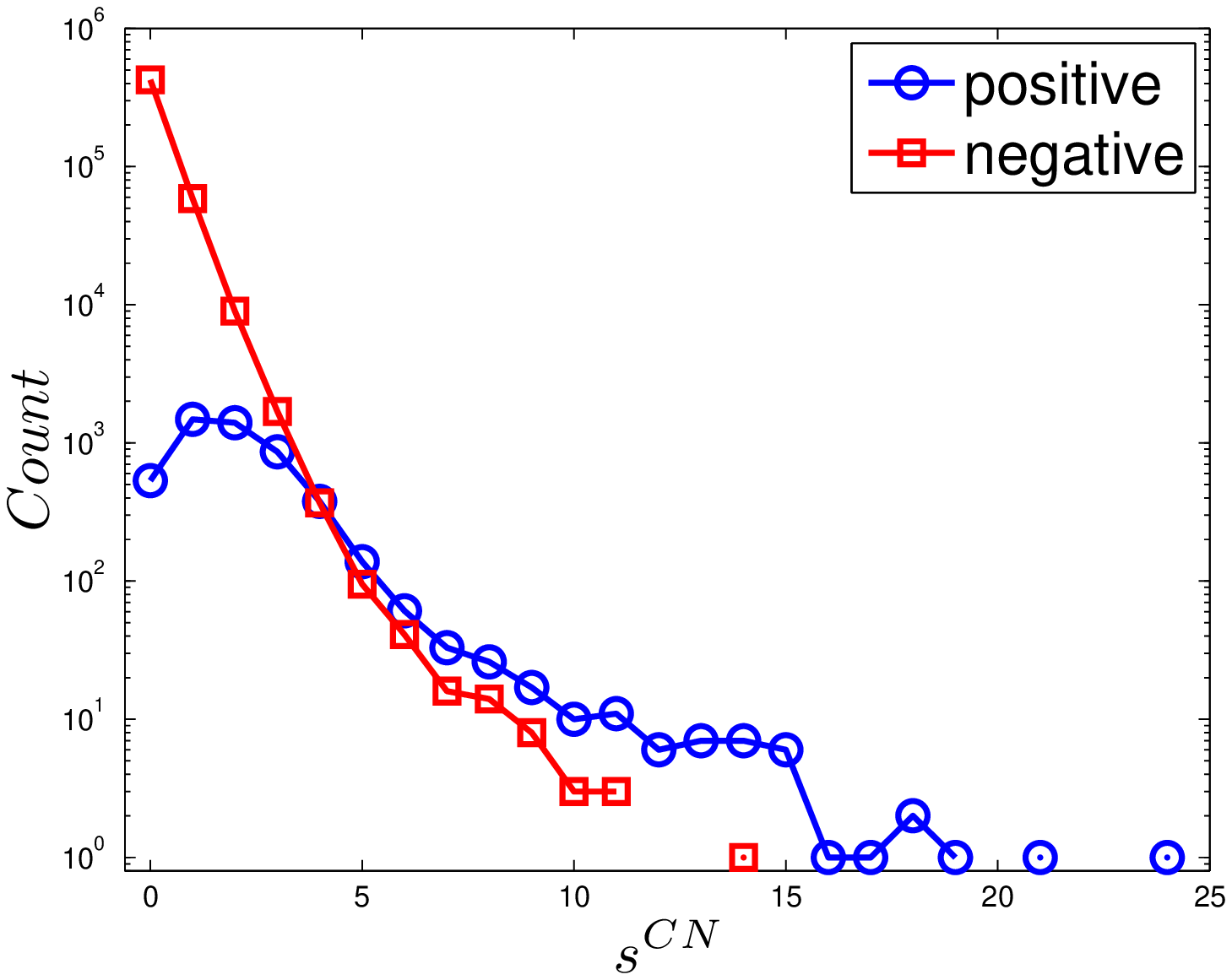,width=5.8cm}\label{fig:ba_cn_c3}}
 \subfigure[\scriptsize \texttt{CN}, $C=0.5,$]{\epsfig{file=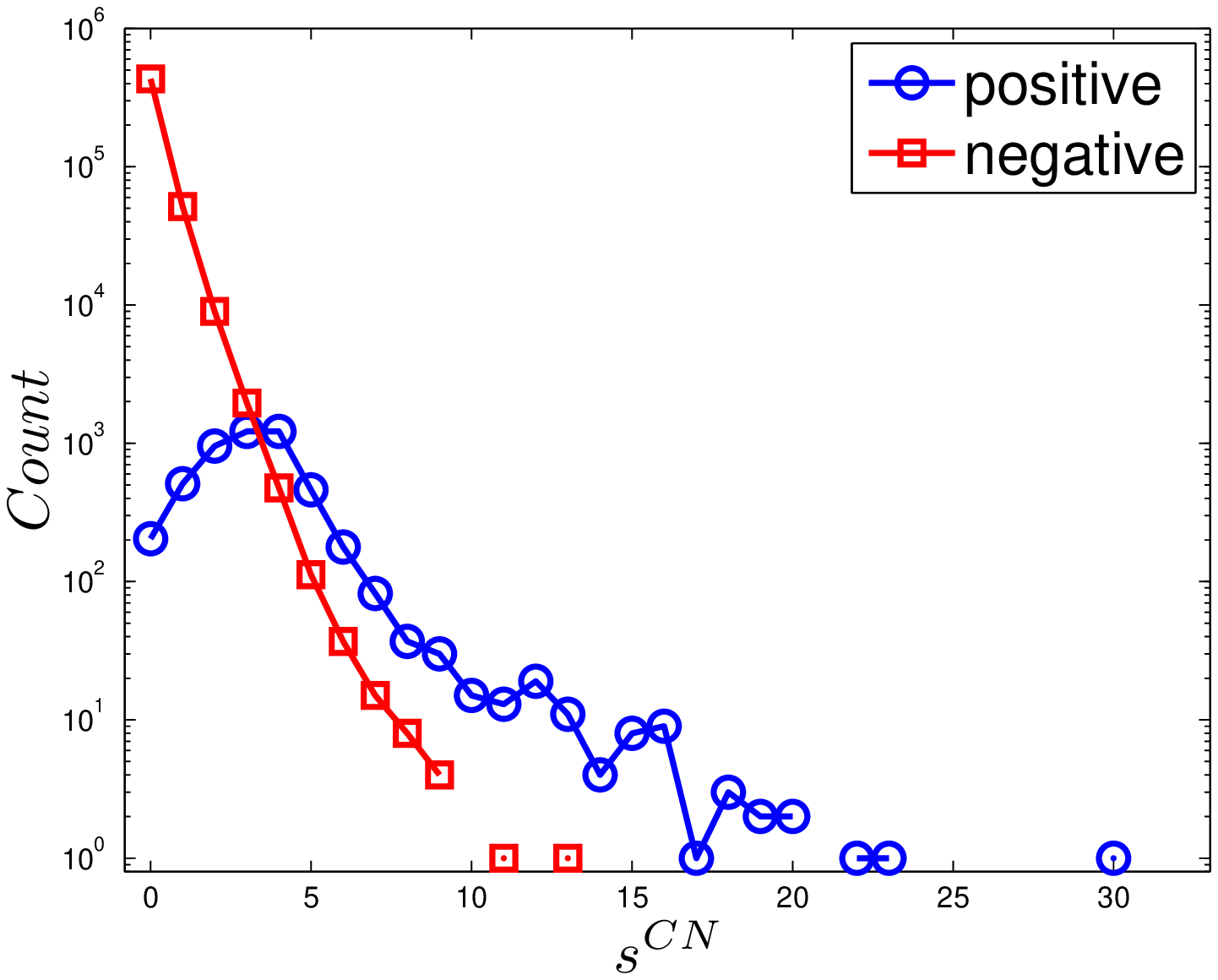,width=5.8cm}\label{fig:ba_cn_c5}}
 \subfigure[\scriptsize \texttt{RA}, $C=0.039$]{\epsfig{file=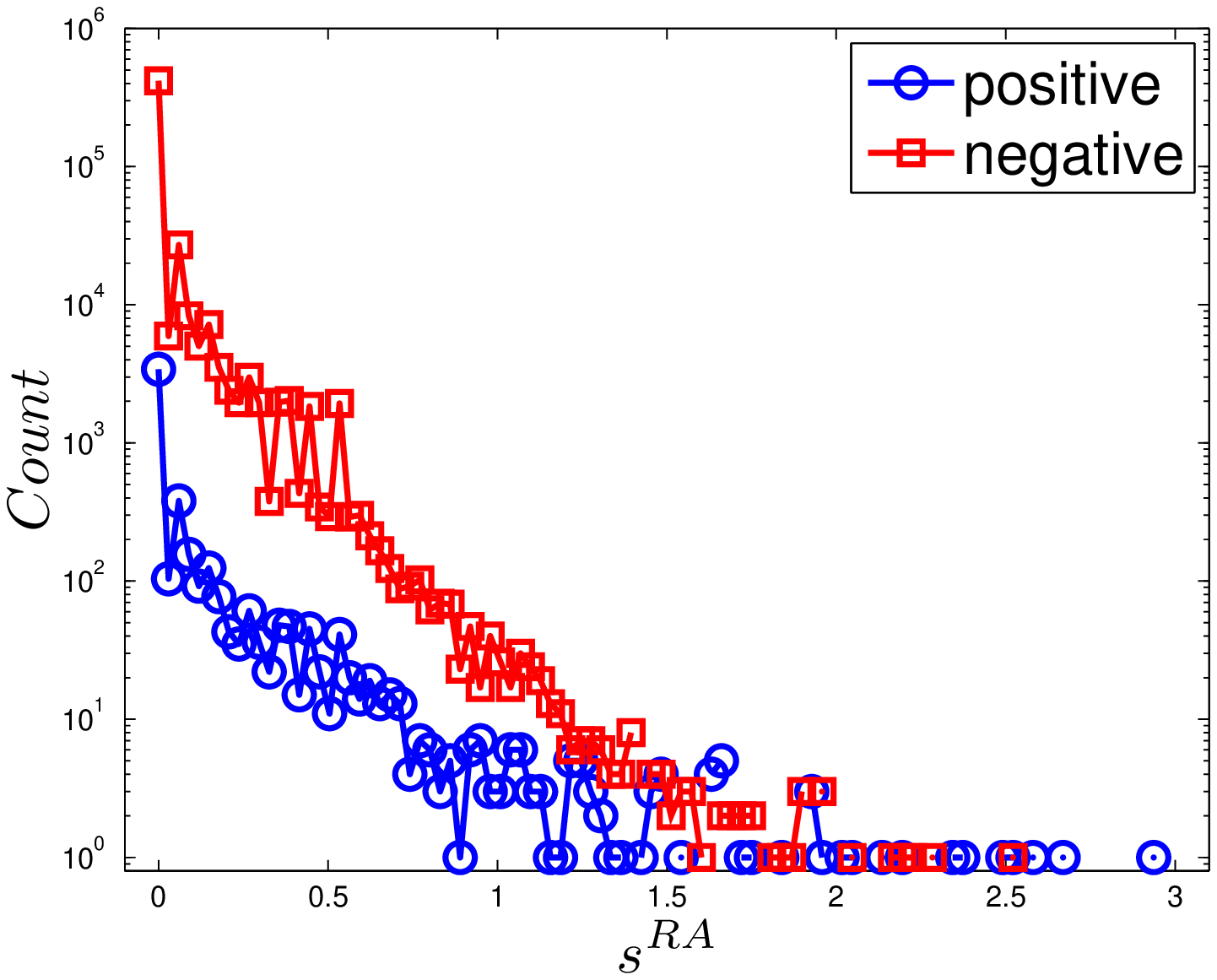,width=5.8cm}\label{fig:ba_ra_origin}}
 \subfigure[\scriptsize \texttt{RA}, $C=0.3$]{\epsfig{file=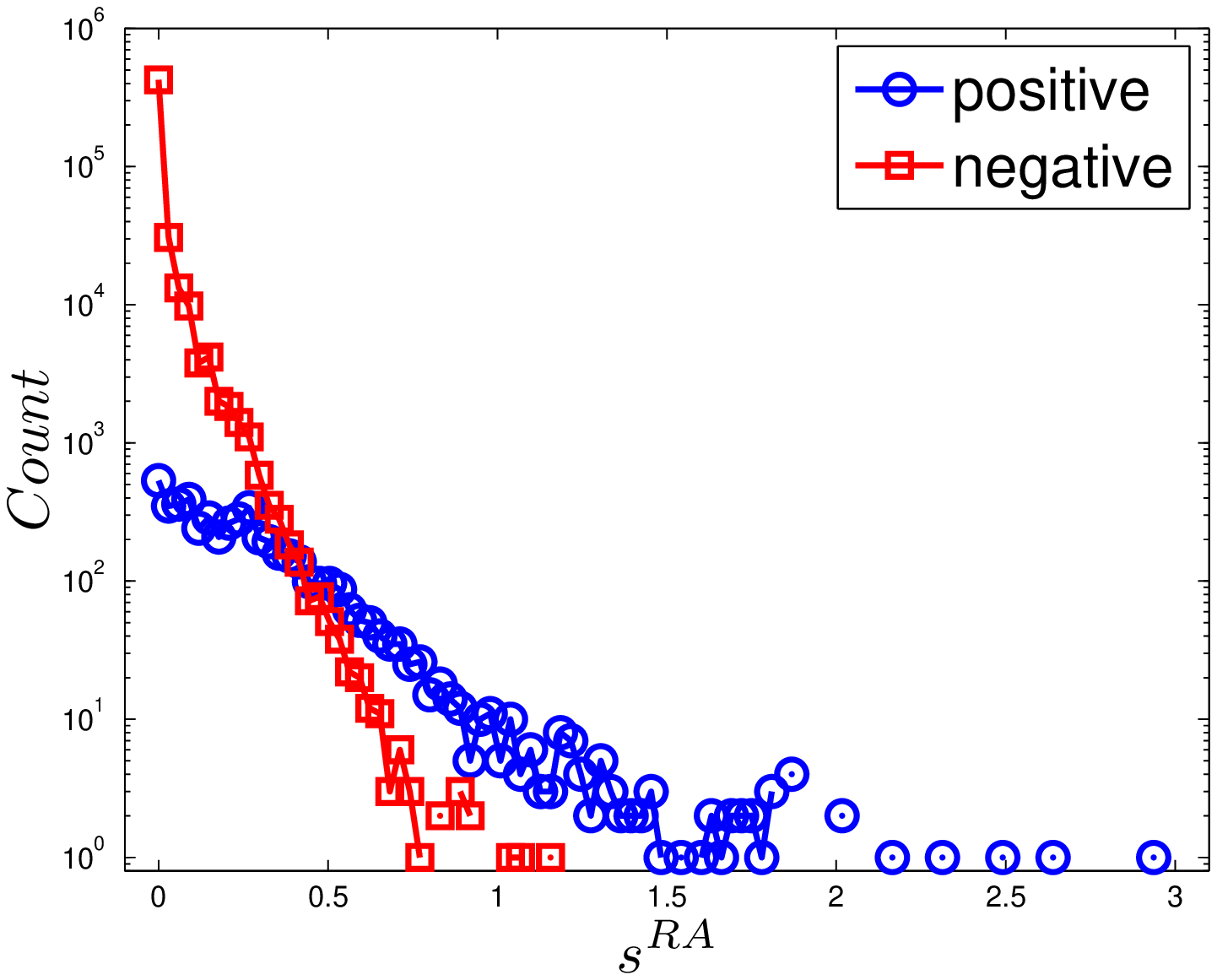,width=5.8cm}\label{fig:ba_ra_c3}}
 \subfigure[\scriptsize \texttt{RA}, $C=0.5$]{\epsfig{file=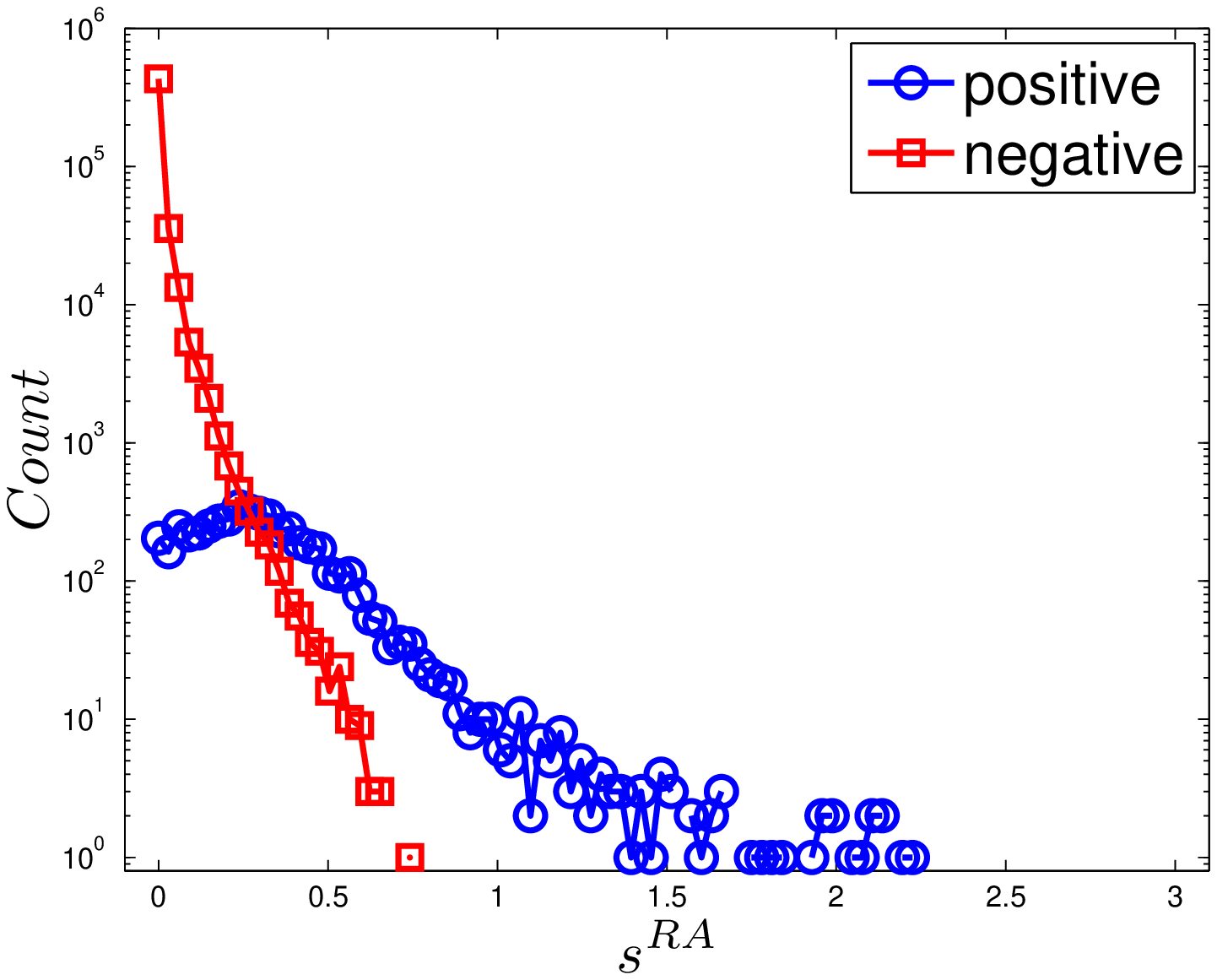,width=5.8cm}\label{fig:ba_ra_c5}}
\label{fig:counts} \caption{Relative distribution of scores for
positive and negative instances as $C$ increases.}
\end{figure*}

As shown in Fig.~\ref{fig:ba_cn_origin}, Fig.~\ref{fig:ba_cn_c3} and
Fig.~\ref{fig:ba_cn_c5}, for \texttt{CN}, we can see a clear
separation between the distributions of $s^{CN}$ on positive and
negative pairs while $C$ of the network increases from 0.039 to 0.5.
This trend can also be observed with \texttt{RA}, as shown in
Fig.~\ref{fig:ba_ra_origin}, Fig.~\ref{fig:ba_ra_c3}, and
Fig.~\ref{fig:ba_ra_c5}. As clustering of the network increases, node pairs with
higher scores are more likely positive instances. Remember that in the link prediction
process, a higher score means a higher probability that
a link will emerge or more likely be missed, so we can conclude that
these prediction methods are more effective in networks with greater clusterings.
In summary, the increment of clustering
improves the capability of these methods for distinguishing the
positive and negative node pairs, which leads to a higher prediction
precision as the experiment shows.

\section{Conclusion}
\label{sec:conslusion}

Link prediction is an open problem in the complex network, which
attracts wide attention in recent years. Plenty of methods
have been presented, some of which are solely based on the structure
while some of which take other features of the network into account. However,
in the real-world, the simpleness and freedom of need for rich
attributes are necessary to the practical methods. For this reason,
we mainly investigate the relationship between six structural
approaches and the clustering of networks. It is interesting that we
find the performance of these methods improves tremendously as the
clustering increases both on synthetic and real-world networks. We also
give this a brief explanation through the extent of
distinguishment between the distribution of positive and negative
instances caused by the variation of clustering. Our finding also
sheds light on the problem of how to choose a simple
but effective method when we meet real networks with various
clusterings. We conjecture that for the sparse network with lower
clustering, \texttt{SRW} is the best choice, while for the network
which is dense and highly clustered, the best choice is \texttt{RA}.

\section{Acknowledgement}
\label{sec:acknowledgement}

This work was supported by the fund of the State Key Laboratory of
Software Development Environment (SKLSDE-2008ZX-03).
%

%

\end{document}